\documentstyle[12pt,aaspp,psfig]{article}
\singlespace
\tighten
\onecolumn
\received{}
\accepted{}
\tighten
\begin{document}

%%%%%%%%%%%%%%%%%%%%%%%%%%%%   EM's DEFINITIONS  %%%%%%%%%%%%%%%%%%%%%%%%%
\def\sun{\hbox{$\odot$}}
\def\earth{\hbox{$\oplus$}}
\def\chbigfont{ \font\preloaded=cmr9 scaled \magstep0 \bf \preloaded}
\def\lesssim{\mathrel{\hbox{\rlap{\hbox{\lower4pt\hbox{$\sim$}}}\hbox{$<$}}}}
\def\gtrsim{\mathrel{\hbox{\rlap{\hbox{\lower4pt\hbox{$\sim$}}}\hbox{$>$}}}}
\def\slantfrac#1#2{\hbox{$\,^#1\!/_#2$}}
\def\onehalf{\slantfrac{1}{2}}
\def\onethird{\slantfrac{1}{3}}
\def\twothirds{\slantfrac{2}{3}}
\def\onequarter{\slantfrac{1}{4}}
\def \threequarters{\slantfrac{3}{4}}
\def \Point{{\bf $\bullet$ } }
\def \be{\begin{equation}}
\def \ee{\end{equation}}
\def \and{\& }
\def \acknowledgments{\vskip 3ex plus .8ex minus .4ex}
\let \acknowledgements=\acknowledgments			
%%    U N I T S
\def \units{\hspace{1cm}}
\def \erg {~{\rm erg}}
\def \cm{~{\rm cm}}
\def \solarmass{~\! M_\odot}
\def \Msun{~ M_\odot}
\def \Lsun{~ L_\odot}
\def \pc{~{\rm pc}}
\def \kpc{~{\rm kpc}}
\def \Mpc{~{\rm Mpc}}
\def \mas{~{\rm mas}}
\def \km{~{\rm km}}
\def \sec{~{\rm s}}
\def \yr{~{\rm yr}}
\def \year{~{\rm yr}}
\def \persecond{~{\rm s}^{-1}}
\def \persec{~{\rm s}^{-1}}
\def \peryear{~{\rm yr}^{-1}}
\def \arcmin{\hbox{$^\prime$}}
\def \arcsec{\hbox{$^{\prime\prime}$}}
\def \micron{\hbox{$\mu$m}}
\def \deg{\hbox{$^\circ$}}
\def \ster{~{\rm ster} }
\def \GHz{~{\rm GHz} }
\def \Kelvin{~\!{\rm K} }
\def \Jy{~{\rm Jy} }
\def \period{\hspace{1cm} . }
\def \coma{\hspace{1cm} , }

\def \bfv{{\bf v}}
\def \spa{~\!}
\def \los{l.o.s. }
\def \etal{{\it et~al.~}}
\def \etalb{{\it et}$\!$ {\it al.}$\!$}
\def \eg{{\it e.g.~}}
\def \ie{{\it i.e.~}}
\def \cf{{\it c.f.~}}

%%%%%%%%%%%%%%%%%%%%%%%%%%%%% END OF EM's DEFINITION  %%%%%%%%%%%%%%%%
\tiny
.
\normalsize

\vspace{0.7truein}
 
\title{A Test For An Intervening Stellar Population As The Origin\\
of Microlensing Events Toward the Large Magellanic Cloud}

\vspace{0.8in}
\author{Eyal Maoz}
% \author{Eyal Maoz$^{1}$ \, and \, Charles Alcock$^{2}$ }
\vspace{0.4in}
\affil{Astronomy Department, University of California, 
Berkeley, CA 94720}

\vspace{0.2in}
% \affil{$^{2}$Lawrence Livermore National Laboratory, Livermore, CA 94551}

\vspace{2.4in}
\centerline{$\dagger$ \it Submitted to the Astrophysical Journal (Letters)}

\def\anewpage{\newpage}
\newpage
%%%%%%%%%%%%%%%%%%%%% ABSTRACT %%%%%%%%%%%%%%%%%%%%%%%%%%%%%%%%%%%%
\begin{abstract}
Zaritsky and Lin (1997) have recently presented observations which can
be interpreted as evidence for an intervening stellar population along the
line-of-sight to the LMC, at a distance of about $34\kpc$. 
The evidence is based on detected populations of red-clump and lower
main-sequence stars which appear brighter than the principal distributions
by about 0.9 magnitude, and which have similar fractional surface densities of
$\sim\!6\%$.  ~ZL97 suggest that the intervening population may belong to a 
low surface-brightness dwarf galaxy, or to the tidal debris tail of a disrupted 
galaxy.  They derive an optical depth for gravitational lensing 
through the intervening system which is consistent with
the observed value, thus suggesting that MACHOs 
comprise at most a small fraction of the Galactic dark halo.

Such interpretation of the microlensing events toward the LMC has major
implications for the nature of the dark matter, and thus should be tested in
every conceivable way. 
We suggest a test which is based on the microlensing data themselves: 
if the observed events arise in the ZL97 intervening population, 
then all the lenses are at nearly the same distance, and
have nearly the same transverse velocity. 
Thus, the predicted distribution of event durations should directly reflect
the stellar mass function in the intervening population, which is probably 
not significantly different from that in the Galactic disk.  

We find that the first eight MACHO events toward the LMC are as consistent
with a standard halo distribution of $0.5\solarmass$ MACHOs (best fit), as they
are with an intervening population having a transverse speed
of $\sim\!240\km\persec$ and a Miller-Scalo stellar IMF.
~Monte Carlo simulations reveal that in order
to completely rule out either the standard halo distribution of 
MACHOs or the ZL97 intervening population, about 40 
events will be required if the IMF is rising 
below the H-burning mass limit, and $\approx\!140$ if it is falling.
If the intervening population hypothesis is correct then: (a) the 
currently observed range of event durations directly implies a lens mass range
of $\approx 0.06\hbox{--}0.8\solarmass$, 
which brackets the mass of $\approx\!0.2\solarmass$
where many investigations claim that the stellar mass function peaks,
and (b) the event distribution provides
valuable information on the low-mass end of the stellar IMF. 

\end{abstract}


\newpage 
\def \dLMC{d_{LMC}}
\def \vlos{v_{los}}
\def \MACHOs{MACHOs}
\def \MACHO{MACHO}
%%%%%%%%%%%%%%%%%%%%%%%%%%%%%%%%%%%%%%%%%%%%%%%%%%%%%%%%%
\section{INTRODUCTION}

Ground based observations of stars in the Large Magellanic Cloud (LMC)
have recently revealed a population of red-clump stars that appear 
brighter by about 0.9 magnitude than the peak of the principal 
red clump distribution (Zaritsky \& Lin 1997, hereafter ZL97). 
After considering
and rejecting various possible explanations, ZL97 argue that these red-clump
stars belong to a spatially localized intervening population of stars along 
the line-of-sight to the LMC, at a distance of about $34\kpc$. 
ZL97 present corroborating evidence for this interpretation from 
HST observations of LMC stars (Holtzman \etal 1997). These data 
show a faint trace of a secondary lower main sequence population which is 
shifted by 0.9 mag and has a fractional surface density similar to that of
the brighter red-clump stars ($\sim\!6\%$).
Using the derived distance and estimated projected density of the 
proposed intervening population, ZL97 derived an 
optical depth for gravitational lensing which is within a factor of two 
of the observed value, thus suggesting that MACHOs with masses in the range of 
$\sim\!0.1\hbox{--}1\solarmass$
comprise at most a small fraction of the Galactic dark halo.

These findings seem to support an earlier suggestion (Zhao 1997)
that the observed gravitational microlensing events toward the LMC may be
due to a low surface-brightness dwarf galaxy or the tidal 
debris of a disrupted system at the foreground of the LMC.
Such interpretation of the microlensing data would have major implications for
the nature of the dark matter, and thus should be tested in 
every conceivable way.
Several potential tests have 
already been suggested: (a) looking for spatial correlations between 
the red-clump and the brighter red-clump stars in fields farther from the LMC 
center, as attempted by ZL97 who found suggestive support for the 
foreground population hypothesis; (b) analyzing variable stars in microlensing 
databases, as performed by Alcock \etal (1997b) who excluded the possibility
of a dwarf galaxy along the l.o.s.~to the LMC out to a distance of $30\kpc$;
examining surface-brightness maps within large areas around the LMC
(Gould 1997); and (d) looking for 
differences in the radial velocity distributions of the principal 
red-clump and brighter red-clump stars, as suggested by ZL97. 
 
In the present paper, we suggest a method utilizing 
the observed microlensing events
themselves to discriminate between the possibilities that they arise
in a standard halo distribution of MACHOs, or in 
an intervening stellar population at $\sim\!34\kpc$.  We demonstrate 
that the distribution of event durations should be different in these 
two cases, and examine the likelihood 
that the observed events are drawn from either population (\S2).
Using Monte Carlo simulations, we estimate 
the number of events required to safely rule out either hypothesis.
In \S3 we discuss possible
implications of the results to the nature of the lensing
population, and to the low-mass end of the stellar IMF.

\anewpage
%%%%%%%%%%%%%%%%%%%%%%%%%%%%%%%%%%%%%%%%%%%%%%%%%%%%
\section{DISTRIBUTION OF MICROLENSING EVENT DURATIONS}
%%%%%%%%%%%%%%%%%%%%%%%%%%%%%%%%%%%%%%%%%%%%%%%%%%%%

The characteristic timescale of a gravitational lensing event towards the LMC
is given by
\be \hat{t} ~=~ {2\spa r_E \over v_t} ~\equiv~ {4
\left[{G m x (1- x) \dLMC}\right]^{1/2} \over v_t \spa c}, 
\label{def_t_hat} \ee
where $r_E$ is the Einstein radius, $x\!\equiv\! d/\dLMC$,
$d$ is the distance to a lensing object of mass $m$, $\dLMC$ is 
the LMC distance ($\simeq\!50\kpc$; Feast \& Walker 1987), 
and $v_t$ is the transverse
velocity of the Einstein ring relative to the line-of-sight to the source.
Therefore, the shape of the distribution of event durations depends on the 
density distribution of lenses along the l.o.s., $\rho(x)$, 
their velocity distribution, and their mass function.  
These ingredient distributions are very different in the two cases between 
which we wish to discriminate.
In the case of a dark halo distribution of \MACHOs, $\rho(x)$ is reasonably
well determined, and the distribution of transverse velocities can be 
derived by assuming an isothermal velocity distribution, taking 
into account the Solar and LMC motions. The MACHO mass function is 
a priori unknown. 

By contrast, in the case of the intervening population proposed
by ZL97, all the lenses are at nearly the same distance, 
$d_{int}\!\approx\!34 \kpc$, and have similar transverse velocities.
The reason for the latter is the following: 
the angular size of the intervening system must not be much smaller 
than that of the LMC since otherwise microlensed stars would have appeared 
clustered in a limited area of the LMC (Maoz 1994). In addition, the mean
surface mass density of the intervening population should not exceed 
\be \Sigma_{int} \spa \lesssim \spa {c^2 \spa \tau_{obs} ~\!\dLMC\over 
 4\pi G d_{int}(\dLMC - d_{int})} ~\simeq ~44 
 \spa \left({\tau_{obs} \over 2.9\!\times\!10^{-7}}\right) ~ 
\solarmass \pc^{-2} \ee
in order to be consistent with the inferred optical depth for microlensing 
toward the LMC (Alcock \etal 1997; hereafter A97).
If the intervening system is bound, the dispersion in transverse 
velocities is $\sigma_{int,\perp}\!\sim\!(GM_{int}/R_{int})^{1/2} \!\sim\!
(\pi G \Sigma_{int} R_{int})^{1/2}$, where $R_{int}$ is the system's
characteristic radius. This yields $\sigma_{int,\perp}\!\lesssim\!24 \spa
(R_{int}/{\rm kpc})^{1/2} \km\persec$, which is very likely to to be
negligible relative to the motion of the entire intervening system.
In the case of an unbound or a tidally 
disrupted system, the total solid angle of the intervening structure must
be much smaller than $4\pi$ since otherwise its total mass 
would be $\sim\!4\pi d_{int}^2\Sigma_{int}=6\!\times\!10^{11} \solarmass$,
which exceeds the entire mass of the dark halo.  Thus, if the
intervening population is 
to be relatively confined in space, its velocity dispersion  
must be much lower than the circular velocity in the halo, 
$v_c\!\simeq\!220\km\persec$
(any possible large-scale velocity gradient along 
a tidal tail would not introduce much of a dispersion in the velocities
of those intervening stars within the small solid angle of the LMC).
We conclude that the distribution of event durations should directly reflect
the mass function of the intervening population. Again, in contrast to the
a priori completely unknown MACHO mass function, it is reasonable to 
expect that the stellar 
mass function in the intervening population is not
significantly different from that in the Galactic disk.

\anewpage
%%%%%%%%%%%%%%%%%%%%%%%%%%%%%%%%%%%%%
\subsection{Localized Intervening Stellar Population}
%%%%%%%%%%%%%%%%%%%%%%%%%%%%%%%%%%%%%
\def\mtil{\tilde{m}}
\def\tsun{t_\odot}
\def\xint{x_{int}}

In the case of the intervening stellar population proposed by ZL97, we can
approximate the lens distribution along the l.o.s to the LMC by 
$\delta(x\!-\!\xint)$, where $\xint\!\simeq\!0.68$, and the transverse velocity
distribution of the stars by $\delta(v-v_t)$, where 
\be  v_t ~=~ \left|{ \bfv_{int,\perp} - \bfv_{los}(x_{int}) }\right| \coma
\label{def_vt} \ee
$\bfv_{int,\perp}$ is the transverse velocity vector of the
intervening system, and $\bfv_{los}(x)$ is the transverse velocity of
the line-of-sight to the source, at the distance of a lens, given by
\be \bfv_{los}(x) ~=~  {(1-x) \bfv_{\odot,\perp} ~+~ x \bfv_{LMC,\perp} } \coma \label{def_vlos} \ee
where $\bfv_{\odot,\perp}$ and $\bfv_{LMC,\perp}$ are the transverse velocity
vectors of the Sun and LMC, respectively.
In a Cartesian coordinate system where $v_x$, $v_y$ and $v_z$ are in the
directions of the Galactic center, the Solar motion, and the Galactic North
pole, respectively, the LMC motion with respect to the Galaxy is
$\bfv_{LMC}\!=\!(60\pm59,-155\pm25,144\pm51) \km\persec$ (Jones, Klemola \&
Lin 1994), the Solar motion with respect to the
Galaxy is $\bfv_\odot\!=\!-\bfv_{LSR}\!+\!\bfv_c = (9,231,16) \km\persec$, and
the unit vector in the LMC direction is $\hat{n}_{LMC}\!=\!
(0.146,-0.826,-0.545)$.

The differential event rate is 
\be d\Gamma ~=~2\spa\Psi(\mtil) \spa \mtil^{1/2} \spa v_t \spa 
R_{E}(M_\odot,x_{int}) \spa d\mtil, \label{def_dGdm} \ee
where $\mtil\!\equiv\!m/M_\odot$, $R_E$ is defined in Eq.~(\ref{def_t_hat}),
and $\Psi(\mtil)$ is the present day
mass function of the intervening stellar population, normalized
such that $\int{\Psi(\mtil) m \spa d\mtil} \!=\! \Sigma_{int}$.
Although the universality of the stellar mass function is yet to be proven,
it is reasonable to assume that $\Psi(\mtil)$ is not significantly different 
from the stellar mass function in the Galactic disk. 
 
The understanding of the stellar mass function has advanced
substantially since the concept of the IMF was introduced by Salpeter (1955), 
but at the low-mass end ($\simeq\!0.08\hbox{--}0.2\solarmass$), 
the results of various workers have been discrepant
(\eg see Miller \& Scalo 1979; Simons \& Becklin 1992;
Kroupa, Tout \& Gilmore 1993; Comeron \etal 1993; Tinney 1993). 
On one hand, there are claims for a turnover in the mass function at $\sim\!
0.2\solarmass$, supported by observations of star forming regions
(\eg Hillenbrand 1997) 
and by possible evidence for a falling IMF at 
masses below the H-burning limit ($\simeq\!0.08\solarmass)$ from observations 
of proto-brown dwarf clumps  (Pound \& Blitz 1995), 
thus suggesting that brown dwarfs exist in such 
small numbers as to make a negligible contribution to the mass density 
of the Galactic disk.
At the other extreme, there are claims for a steeply rising mass function
all the way down to the H-burning mass limit 
(\eg Mera, Chabrier, \& Baraffe 1996),
implying a substantial amount of brown dwarfs in the Galactic disk (see also
Tinney 1993).

The shape of the low-mass end of the IMF is likely to be better 
determined in the next few years (\eg due to deep IR searches for brown 
dwarfs), but at present 
we cannot rule out either possibility.  Thus, we shall explore two 
mass functions which differ significantly at low masses: 
(a) the Miller and Scalo model (1979; hereafter MS79) where the IMF 
is fit by a lognormal distribution with a slope 
$d\log\Psi/d\log\mtil\!=\! -(1\!+\!\log\mtil)$, \ie
$\Psi(\mtil)\spa d\mtil \propto \mtil^{-(0.5\log\mtil+2)}d\mtil$ (turns over
at $0.1\solarmass$), and (b) the three-component power-law fit 
of Kroupa, Tout, and Gilmore (1993; hereafter KTG93), which is monotonically
rising down to the H-burning limit, extended here down to a Jupiter mass
\begin{eqnarray}
 \Psi(\mtil) ~\propto ~ \left\{  \begin{array}{lr}
3.5 ~\mtil^{-1.3}   &  \hspace{0.2cm}  0.001 < \mtil < 0.5  \\
1.9 ~\mtil^{-2.2}   &  \hspace{0.2cm}  0.5 < \mtil < 1.0  \\
1.9 ~\mtil^{-2.7}   &  \hspace{0.2cm}  1.0 < \mtil < \infty
                             \end{array}
\right.  \period \end{eqnarray}
The cutoff mass is introduced because the IMF cannot rise indefinitely,
but the choice of a Jupiter mass is arbitrary. We 
note that the results will not be sensitive to the exact cutoff mass since 
the detection efficiency for very short events is low.

Let us define $\tsun$ as the crossing time of an Einstein ring due to one solar 
mass object at a distance of the intervening system, 
\be \tsun \equiv {2 \spa r_{E}(M_\odot,x_{int}) \over v_t} = ~ 166 
\left({v_t \over 200 \km\persec}\right)^{-1} \hspace{0.6cm}{\rm days,} 
\label{def_tsun}  \ee
where $v_t$ is defined in Eq.~(\ref{def_vt}).
Transforming $d\Gamma/d\mtil$ (Eq.~[\ref{def_dGdm}]) into $d\Gamma/d\hat{t}$, 
we obtain for the MS79 mass function
\be {d\Gamma \over d\hat{t}} ~\propto ~ 
\left({\hat{t}/ \tsun}\right)^{-2[\log(\hat{t}/\tsun) +1]} 
\label{def_MS79} \coma \ee
and for the extended KTG93 mass function
\begin{eqnarray}
 {d\Gamma \over d\hat{t}} ~\propto ~ \left\{  \begin{array}{lr}
3.5 ~(\hat{t}/\tsun)^{-0.6}   &  \hspace{0.4cm} 0.03 < \hat{t}/\tsun < 0.707\\
1.9 ~(\hat{t}/\tsun)^{-2.4}   &  \hspace{0.2cm}   0.707 < \hat{t}/\tsun < 1 \\
1.9 ~(\hat{t}/\tsun)^{-3.4}   &  \hspace{0.3cm}  1 < \hat{t}/\tsun < \infty
                             \end{array}
\right.  \period \end{eqnarray}
Figure 1-a presents examples of both event distributions in the case
of $v_t\!=\!100$ and $300\km\persec$.

%%%%%%%%%%%%%%%%%%%%%%%%%%%%%%%%%%%%%
\subsection{Halo Distribution of MACHOs}
%%%%%%%%%%%%%%%%%%%%%%%%%%%%%%%%%%%%%
The event distribution in the case of a dark halo which consists of single mass 
MACHOs with a Maxwellian velocity distribution,
taking the Solar and LMC motions into account, is (Alcock \etal 1996)
\be {d\Gamma \over d\hat{t}} ~\propto~ {1\over m \hat{t}^4 }
\int_0^1{\!dx \spa \rho(x) \spa r_E^{4}(x,m) 
\spa e^{-Q(x,m)} e^{-[\vlos(x)/v_c]^2}
I_0[P(x,m)]  } \label{halo_rate}\ee
where $Q(x,m)\!\equiv\!4 r_E^{2}(x,m)/(\hat{t}^2 v_c^{2})$,
$P(x,m)\!\equiv\!4 r_E(x,m) \vlos(x) /(\hat{t} v_c^2)$,
$I_0$ is the modified Bessel function, and $\vlos(x)\!=\!|\bfv_{los}(x)|$ is
defined in Eq.~(\ref{def_vlos}).
 
We shall consider here only a ``standard halo'' dark matter distribution of 
equal mass MACHOs (A97 present two examples of event distributions
due to power-law mass functions, which closely resemble those
due to equal mass MACHOs).
Following Griest (1991), we take 
$\rho(r)\!=\!\rho_{0}(a^2\!+\!r_0^2)/(a^2\!+\!r^2)$, where
$r$ is the Galactocentric radius,
$a\!=\!5\kpc$ is the core radius, and $r_0\!=\!8.5\kpc$ is the Sun's distance
from the Galactic center. The density along the l.o.s.~to the LMC is
$\rho(x)\!=\!\rho_{0}A'/(A'+Bx+x^2)$,
where $A'\!\equiv\!(a^2+r_0^2)/\dLMC^2\!=\!0.039$,
and $B\!\equiv\!-2(r_0/\dLMC)\cos b_{LMC}\cos l_{LMC}=-0.054$.
%

%%%%%%%%%%%%%%%%%%%%% 
\section{THE MACHO DATA vs. PREDICTED EVENT DISTRIBUTIONS}
%%%%%%%%%%%%%%%%%%%%% 

Our ultimate goal is to rule out either the ZL97 intervening population
hypothesis or the standard halo of MACHOs hypothesis by showing that it is
extremely unlikely that the observed event durations were drawn from
the predicted event distribution.  In order to do that we need to take the
detection efficiency of the relevant experiment into account.
We shall test here the first eight MACHO events toward the LMC (A97)
against the predicted event distributions, 
$\varepsilon(\hat{t})\spa d\Gamma/d\hat{t}$, where $\varepsilon(\hat{t})$ 
is the detection efficiency for the first two years of MACHO data (Figure 2).
The eight event durations are shown at the bottom of Figure 1-b, along with 
examples of the predicted event distributions.

Figure 3 shows the results of Kolmogorov-Smirnov tests (Press \etal 1992), 
where $\alpha$ is the probability of 
getting a K-S deviation worse than the observed value, namely, the
significance level at which we can reject the null hypothesis that 
the data were drawn from a given distribution. For example, 
we can reject at the 0.1 significance level the hypotheses that the data
were drawn from (a) the ZL97 intervening population with an 
extended KTG93 IMF when $v_t$ is either below $30 \km\persec$ or above
$350\km\persec$, ~(b) the ZL97 intervening population with an MS79 IMF when 
$v_t\!\le\!108\km\persec$, ~and (c) a standard halo distribution of 
single-mass MACHOs with $m\!<\!0.15\solarmass$. 
All remaining possibilities are currently consistent with the data, the more 
so for larger $\alpha$. We see that the most likely MACHO mass in a standard
halo model is $0.50\solarmass$, as concluded by A97, but
the ZL97 intervening population with an MS79 IMF and 
$v_t\!\approx\!240 \km\persec$ is equally consistent with the present data.

Eight events are clearly not enough for discriminating between the various
hypotheses. The MACHO experiment has yielded so far an additional six events
toward the LMC (Charles Alcock, private communication), but these are not yet
publicly available.
We performed Monte Carlo simulations to estimate the number of events 
required to safely rule out either the standard halo or the intervening 
population hypothesis, assuming the other hypothesis is correct (Figure 4). 
We find that a sample of $\simeq\!40$ events will enable to rule out a 
significant fraction of the parameter space with $\ge\!95\%$ confidence level, 
but about $140$ events will be required to completely rule out either 
hypothesis. However, this is the worst case scenario estimate 
for several reasons: 
(a) we assumed a detection efficiency similar to that for the first two years
of MACHO data, but $\varepsilon(\hat{t})$ is likely to improve with time, 
especially at long event durations due to the increasing baseline;
(b) the simulations assume that the true event distributions are those with 
$v_t$ and $m$ that best fit the present data, which need not necessarily be
the case; (c) our understanding of the low-mass end of the stellar IMF 
may improve in the near future due to, for example, deep IR searches 
for brown dwarfs; and finally, (d) the permissible range of $v_t$ could
be narrowed by subsequent investigations, as we discuss below.

The transverse velocity of the ZL97 intervening population, $\bfv_{int,\perp}$,
could generally be anything between zero and the escape velocity from the
Galaxy, though not at equal probabilities:
assuming a probability distribution for $\bfv_{int}$ similar to the
velocity distribution function in an isothermal halo,
$P(\bfv_{int})\spa d^{3}v_{int} =
(e^{-v_{int}^{2}/v_{c}^{2}})/(\pi^{3/2}v_c^3) \spa d^{3}v_{int}$, we find that 
the probability distribution for $v_t$ (defined in Eq.~[\ref{def_vt}]) is
\be P(v_t)\spa dv_t ~=~ (v_t/v_c)\spa \exp{[-(v_t^2 + \beta^{2})/v_c^2]} 
\spa I_{0}(v_t \beta/v_c^2) \spa dv_t \label{prob_vt} \coma 
\label{def_prob_vt}\ee
where $\beta\!\equiv\!v_{los}(x_{int})\!=\!106\km\persec$. It peaks at 
$v_t\!=\!236 \km\persec$, but the distribution is quite broad.
It may be possible to derive the transverse velocity of the intervening 
population by a careful re-examination of the data which led to the 
determination of the LMC's proper motion (Jones \etal 1994). Alternatively,
if the radial velocity of the proposed intervening population is measured, 
then dynamical models may well constrain its transverse motion. 
Theoretical models may also yield predictions for the transverse
velocity. For example, 
Zhao (1997) suggested that the lensing population toward the LMC may be
a part of a tidal debris tail in the Magellanic great circle, in which case
the tidal tail and the LMC are expected to have similar specific angular
momentum. Neglecting the offset of the Sun from the Galactic
center and substituting $\bfv_{int}\!=x_{int}^{-1}\spa\bfv_{LMC}$ in
Eq.~[\ref{def_vt}], we obtain for that scenario $v_t\!\approx\!201\km\persec$.
Therefore, the statistical test proposed in this paper 
may rule out various theories for the origin and nature of the intervening 
population with fewer events than estimated above.

%%%%%%%%%%%%%%%%%%%%% 
\section{DISCUSSION}
%%%%%%%%%%%%%%%%%%%%%

If the observed microlensing events toward the LMC are due to the intervening
stellar population proposed by ZL79, then the currently observed range 
of event durations ($39\hbox{--}143$ days) directly implies a lens mass 
range of $0.06\hbox{--}0.8\spa (v_t/200\km\persec)^2\solarmass$, 
which happens to bracket the mass of $\approx\!0.2\solarmass$ 
where many investigations claim that the stellar IMF peaks 
(see references in \S2.1).
On the other hand, if the observed events are due to a halo distribution of
MACHOs, then the typical MACHO mass is $\sim\!0.5\solarmass$ (A97).
These objects cannot be H-burning stars due to observational 
constraints from deep star counts (Gould, Bahcall \& Flynn 1997), thus leaving
the alternatives of primordial white dwarfs and sub-stellar mass black holes.
It is intriguing that the ZL97 intervening population hypothesis 
could provide a more ``natural explanation'' for the observed range 
of event durations.

We propose a way to test not only the existence of a stellar 
population at the foreground of the LMC, but more importantly, the possibility 
that the observed microlensing events are primarily due to such an 
intervening population.  Since the test relies on the shape of the stellar 
IMF, it could provide invaluable insight into the mass function below
the H-burning mass limit if the ZL97 intervening population hypothesis 
is independently confirmed.  Figure 3 could already tell us that 
if $v_t\!\gtrsim\!230 \km\persec$ then a Miller-Scalo type of an IMF is much 
more likely than a monotonically rising one, and vice-versa if 
$v_t\!\lesssim\!130\km\persec$. 
 
Alcock etal (1996b, hereafter A96) 
placed strict limits on the contribution of low-mass
MACHOs to the dark halo, based on the absence of detected events with 
durations of $0.1\!\lesssim\!\hat{t}\!\lesssim\!20$ days.
If the observed events are due to the ZL97 intervening population, 
this null-detection provides an upper limit to the mass fraction in 
low-mass brown dwarfs. We can estimate the number of expected events due to,
for example, 
$10^{-4}\!\lesssim\!m\!\lesssim\!10^{-2}(v_t/200\km\persec)\solarmass$ objects
by scaling the A96 results:  denoting the fractional contribution of such
objects to the total intervening mass by $f$, 
the expected number of short events is 
$N_{exp}\!\simeq 35f (\tau_{obs}/\tau_{halo})$, 
where 35 is the expected number of events due to
MACHOs in this mass range, assuming they make the entire mass of a standard
dark halo (A96, fig.~10), 
$\tau_{obs}$ is the microlensing optical depth inferred
from the observations, and $\tau_{halo}$ is the predicted optical depth 
through a standard dark halo.  
Since $\tau_{obs}\!\simeq\!0.5\tau_{halo}$ (A97),
we expect $N_{exp}\!\simeq\!17f$ short events
in the case of the intervening population 
hypothesis. For $N_{obs}\!=\!0$, the Poisson $95\%$ confidence level upper 
limit is 3 events, so we obtain an upper limit of $f\!\lesssim\!18\%$ to the 
mass contribution of $10^{-4}\hbox{--}10^{-2}\solarmass$ objects
to the stellar IMF.  This limit will continuously improve with time as long as
no short events are detected.
Constraining the contribution of lower mass objects to
the stellar IMF requires a detailed analysis which is 
beyond the scope of this paper since the detection efficiency of 
``spike'' events does not depend only on $\hat{t}$ (A96). 

\acknowledgments
I thank Charles Alcock for discussion and comments, and Leo Blitz and 
Chris McKee for discussions.

\newpage
%%%%%%%%%%%%%%%%%%%%

\begin{figure}[ht]
\vskip-0.7truein
\textwidth=3in
 \centerline{
\psfig{figure=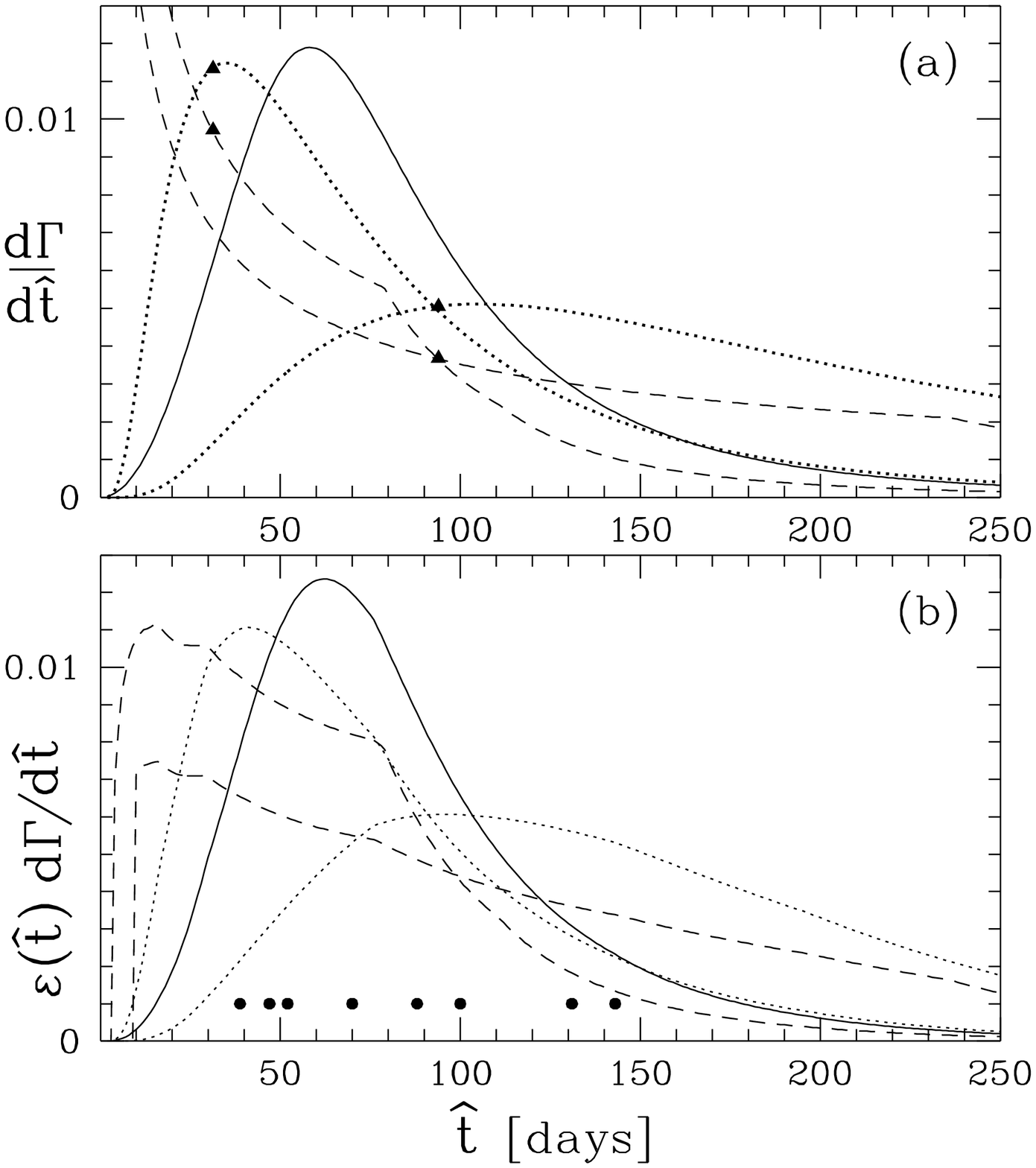,height=6.3in,width=6.4in}}
\caption{ (a) The predicted distribution of event durations toward the LMC
for a standard halo model which consists of $0.5\solarmass$ MACHOs
(solid curve). The dotted (dashed) curves are for the
intervening stellar population suggested by ZL97, assuming an MS79 (extended
KTG) IMFs, respectively. Each set of curves corresponds to transverse
velocities of $v_t\!=\!100$ and $300\km\persec$. The triangles indicate the
event duration due to a $0.08\solarmass$ star (note that
$\hat{t}\!\propto\!m^{1/2}$).
~b)
The predicted distributions for the same parameters, but after taking into
account the detection efficiency for the first two years of MACHO
data (Fig.~2). The points at
the bottom show the durations of the first eight MACHO events (A97).
All distributions are normalized to a unity area within
$2\!\le\!\hat{t}\!\le\!300$ days,
and take the Solar and LMC motions into account. }
\end{figure}

\begin{figure}[ht]
\vskip-0.7truein
\textwidth=3in
 \centerline{
\psfig{figure=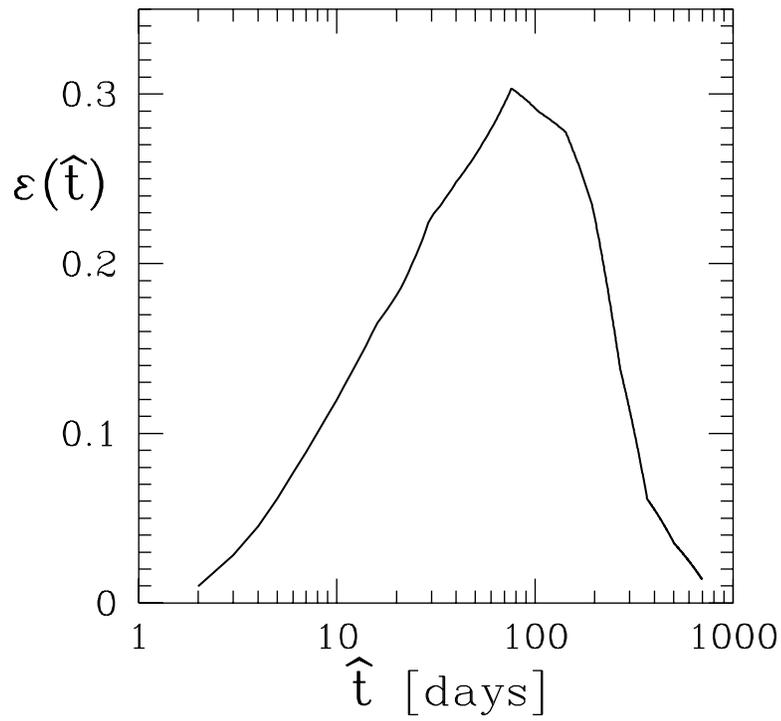,height=6.0in,width=6.0in}}
\caption{The detection efficiency for the first 2-year MACHO data (A97),
kindly provided by Charles Alcock. }
\end{figure}

\begin{figure}[ht]
\vskip-0.7truein
\textwidth=3in
 \centerline{
\psfig{figure=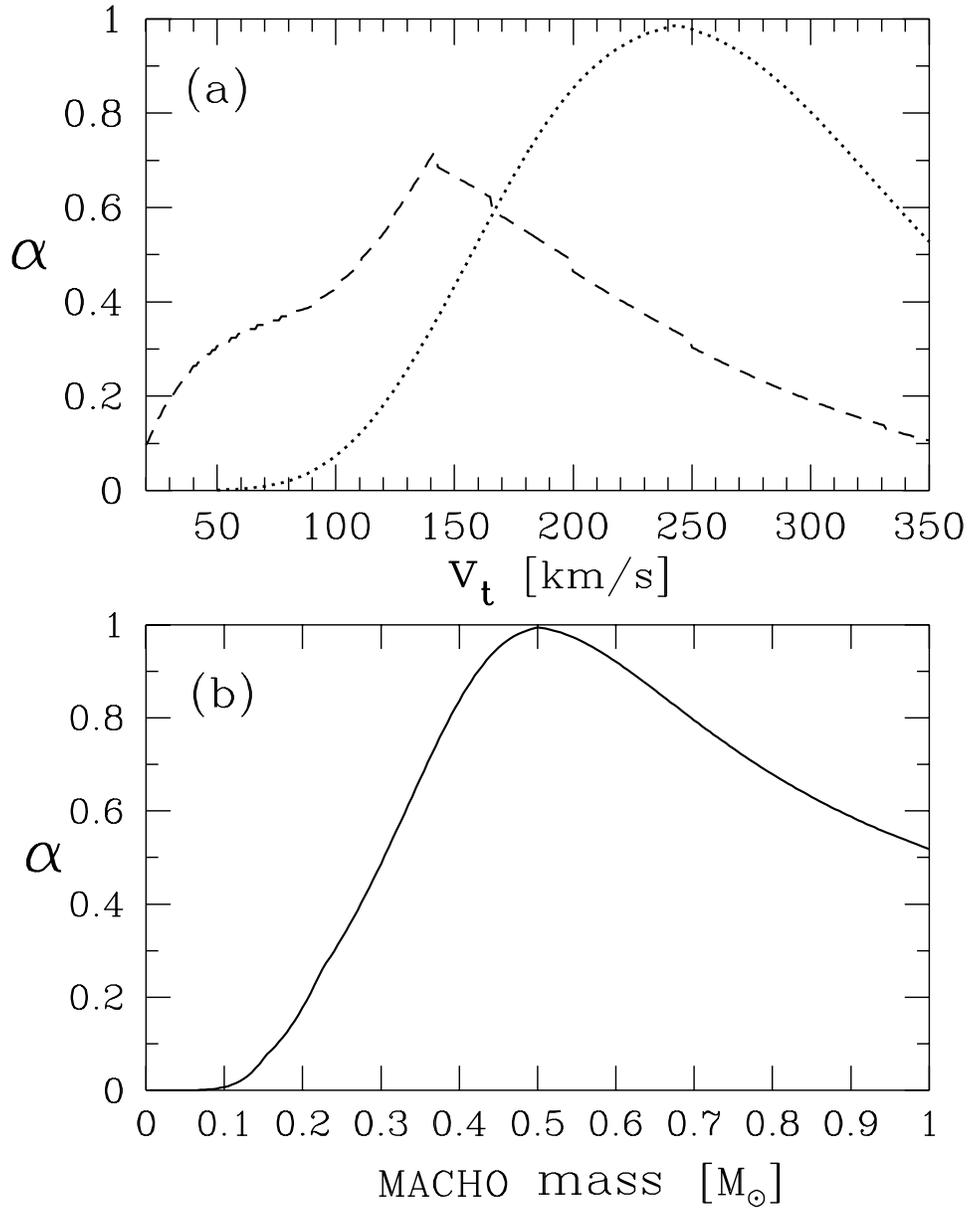,height=6.4in,width=6.4in}}
\caption{The results of K-S tests for the hypotheses that the first eight
MACHO events were drawn from the predicted event distribution of (a) the
ZL97 intervening population with transverse velocity $v_t$ and an MS79 IMF
(dotted); same with an extended KTG93 IMF (dashed), and (b) a standard halo
distribution of MACHOs of mass $m$. The lower $\alpha$, the more inconsistent
the data are with the predicted event distribution (see text). }
\end{figure}

\begin{figure}[ht]
\vskip-0.7truein
\textwidth=3in
 \centerline{
\psfig{figure=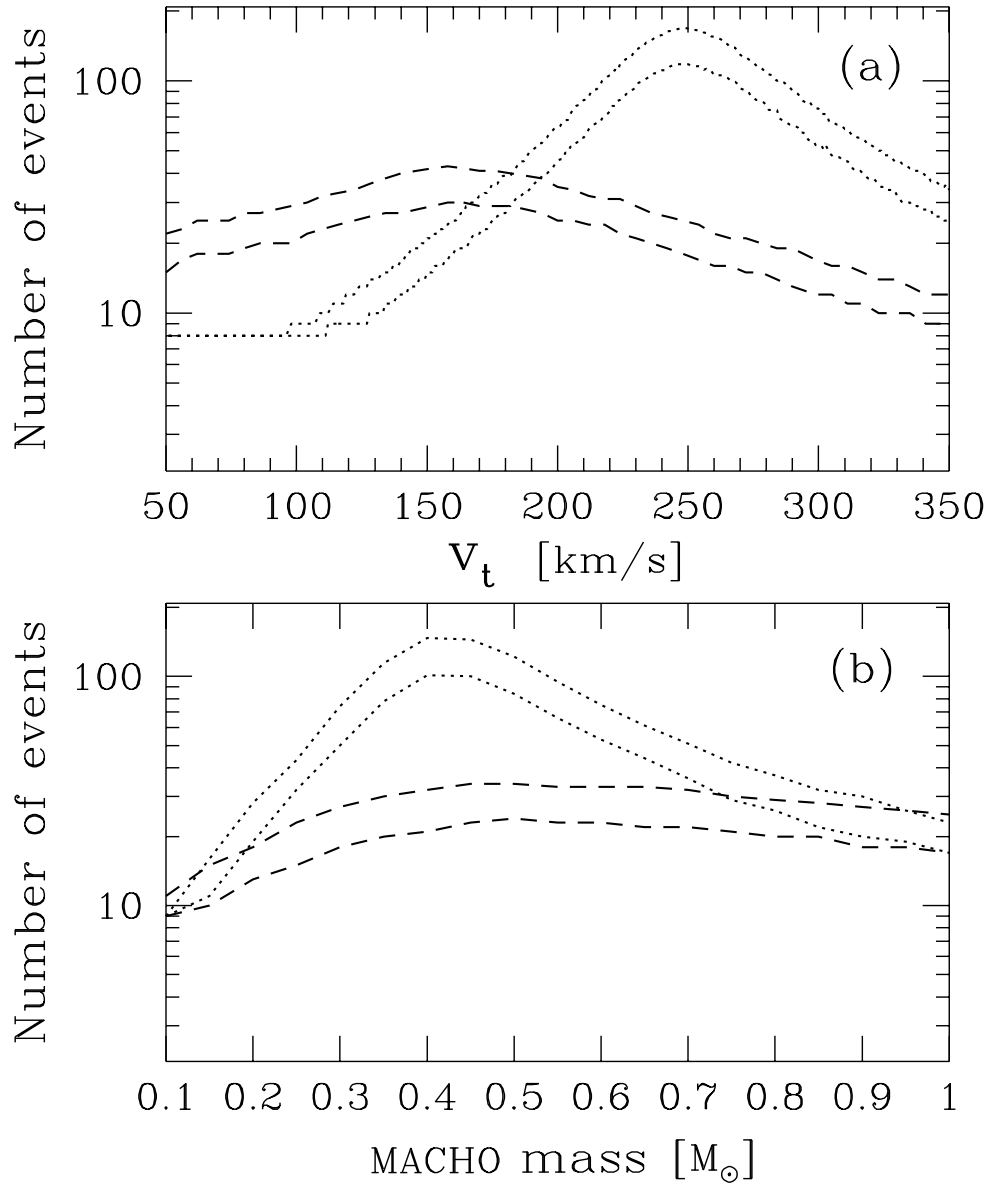,height=6.3in,width=6.4in}}
\caption{ Results of Monte Carlo simulations:  (a) the mean number of events
required to rule out (using a K-S test) a ZL97 intervening population with a
transverse velocity $v_t$, for an MS79 (dotted) and an extended KTG93 (dashed)
IMFs, assuming the events arise in a standard halo distribution of
$0.5\solarmass$ MACHOs (the currently best fit to the MACHO mass).
The upper and lower curves in each pair are for $\alpha\!=\!0.05$ and $0.1$
significance levels, respectively.
~(b) the number of events required to rule out a standard halo of single-mass
MACHOs of mass $m$, assuming the events arise in a ZL97 intervening
population with (1) an MS79 IMF and best fit $v_t$ of $245\km\persec$ (dotted),
and with (2) an extended KTG93 IMF and best fit $v_t$ of $140\km\persec$
(dashed).  See discussion in text. }
\end{figure}
\end{document}